\documentstyle[twocolumn,aps,graphicx]{revtex}

\begin{document}

\twocolumn[\hsize\textwidth\columnwidth\hsize\csname
@twocolumnfalse\endcsname

\title{Why do naked singularities form in gravitational
collapse?}

\author{Pankaj S. Joshi$^*$, Naresh Dadhich$^\dag$ and Roy
Maartens$^\ddag$}

\address{~}

\address{$^*$Tata Institute for Fundamental Research, Mumbai,
India}

\address{$^\dag$Inter-University Centre for Astronomy \&
Astrophysics, Pune, India}

\address{$^\ddag$Institute of Cosmology \& Gravitation, University of Portsmouth,
Portsmouth~PO1~2EG, Britain}

\maketitle

\begin{abstract}

We investigate what are the key physical features that cause the
development of a naked singularity, rather than a black hole, as
the end-state of spherical gravitational collapse. We show that
sufficiently strong shearing effects near the singularity delay
the formation of the apparent horizon. This exposes the
singularity to an external observer, in contrast to a black hole,
which is hidden behind an event horizon due to the early formation
of an apparent horizon.

\end{abstract}


\vskip 1cm]

\section{Introduction}

In the past decade or so, several scenarios have been discovered
where the gravitational collapse of a massive matter cloud results
in the development of a naked singularity~\cite{1}. The final
outcome of gravitational collapse in general relativity is an
issue of great importance and interest from the perspective of
black hole physics as well as its astrophysical implications. When
there is a continual collapse without any final equilibrium,
either a black hole forms when the super-dense regions of matter
are hidden from the outside observer within an event horizon of
gravity, or a naked singularity results as the end product,
depending on the nature of the initial data from which the
collapse develops.

The theoretical and observational properties of a naked
singularity would be quite different from those of a black hole
(see~\cite{jdm} for further discussion of this). Thus it is of
crucial importance to understand what are the key physical
characteristics and dynamical features in collapse that give rise
to a naked singularity, rather than a black hole. While many
models of naked singularity formation within dynamically
developing collapse scenarios have been found and
analyzed~\cite{hm}, not much attention has been given to
understanding this important aspect. We begin here an
investigation of this question.

The main purpose of this paper is to identify the physical process
which exposes the singularity. We find that it is shearing effects
which, if sufficiently strong near the central worldline of the
collapsing cloud, would delay the formation of the apparent
horizon so that the singularity becomes visible and communication
from the very strong gravity regions to outside observers becomes
possible. When the shear is weak (and in the extreme case of no
shear), the collapse necessarily ends in a black hole, because an
early formation of the apparent horizon leads to the singularity
being hidden behind an event horizon.

For spherical gravitational collapse of a massive matter cloud,
the interior metric in comoving coordinates is
\begin{equation}\label{1}
ds^2 = - e^{2\nu(t,r)} dt^2 + e^{2\psi(t,r)} dr^2 + R^2(t,r)
d\Omega^2\,.
\end{equation}
The matter shear is
\begin{equation}\label{17}
\sigma_{ab} = e^{-\nu} \left({\dot R \over R} - \dot
\psi\right)({\textstyle{1\over3}}h_{ab} -n_an_b)\,,
\end{equation}
where $h_{ab}=g_{ab}+u_au_b$ is the induced metric on 3-surfaces
orthogonal to the fluid 4-velocity $u^a$, and $n^a$ is a unit
radial vector.

The initial data for collapse are the values on  $t=t_{\rm i}$ of
the three metric functions, the density, the pressures, and the
mass function that arises from integrating the Einstein equations
(for details see e.g.~\cite{3}),
\begin{equation}\label{4}
F(t_{\rm i},r) = \int \rho(t_{\rm i},r) r^2 dr\,,
\end{equation}
where $4\pi F(t_{\rm i},r_{\rm b})=M$, the total mass of the
collapsing cloud, and where $r>r_{\rm b}$ is a Schwarzschild
spacetime. We use the rescaling freedom in $r$ to set
\begin{equation}\label{3}
R(t_{\rm i},r) = r\,,
\end{equation}
so that the physical area radius $R$ increases monotonically in
$r$, and with $R_{\rm i}'=1$ there are no shell-crossings on the
initial surface. (We will be interested here only in the central
shell-focusing singularity at $R=0,r=0$ which is a gravitationally
strong singularity, as opposed to the shell-crossing ones which
are weak, and through which the spacetime may sometimes be
extended.) The evolution of the density and radial pressure are
given by
\begin{equation}\label{5}
\rho = {F'\over R^2 R'}, \quad p_{\rm r} = {\dot F\over R^2 \dot
R}\,.
\end{equation}
The central singularity at $r=0$, where density and curvature are
infinite, is naked if there are outgoing nonspacelike geodesics
which reach outside observers in the future and terminate at the
singularity in the past. Outgoing radial null geodesics of
Eq.~(\ref{1}) are given by
\begin{equation}\label{6}
{dt\over dr} = e^{\psi - \nu}\,.
\end{equation}

Consider first the case of homogeneous-density collapse,
$\rho=\rho(t)$. Writing $f= e^{-2\psi} R'^2-1$, the Einstein
equations give $f - e^{-2\nu} \dot R^2 = - {F/R}$. Then
Eq.~(\ref{6}) can be written as~\cite{4}
\begin{equation}\label{8}
{dR\over du} = \left( 1 - \sqrt{ {f + {F/ R}\over 1 +f}}\right)
{R'\over \alpha r^{\alpha -1}}\,,
\end{equation}
where $u=r^\alpha$ ($\alpha >1$). If there are outgoing radial
null geodesics terminating in the past at the singularity with a
definite tangent, then at the singularity we have $dR/du > 0$. For
homogeneous density, the entire mass of the cloud collapses to the
singularity simultaneously at the event $(t=t_{\rm s},r=0)$, so
that $F/ R \to \infty$. By Eq.~(\ref{8}), ${dR/ du} \to -\infty$,
so that no radial null geodesics can emerge from the central
singularity. It can be similarly shown that all the later epochs
$t>t_{\rm s}$ are similarly covered.

We have thus shown that {\em for spherical gravitational collapse
with homogeneous density (and arbitrary pressures), the final
outcome is necessarily a black hole}. We note that this conclusion
does not require homogeneity of the pressures $p_{\rm r}$ and
$p_\perp$, and is independent of their behavior. The result
generalizes the well-known Oppenheimer-Snyder result for the
special case of dust, where the homogeneous cloud collapses to
form a black hole always.

An immediate consequence is that {\em if the final outcome of
spherical gravitational collapse is not a black hole, then the
density must be inhomogeneous}. In any physically realistic
scenario, the density will be typically higher at the center, so
that generically collapse is inhomogeneous.

\section{Inhomogeneous dust}

Consider now a collapsing inhomogeneous dust cloud ($p=0$), with
density higher at the center. The metric is
Tolman-Bondi-Lema{\^i}tre, given by Eq.~(\ref{1}) with $\nu=0$ and
$e^{2\psi}=R'^2/(1+f)$, and
\begin{equation}\label{12}
\dot R^2 = f(r) + {F(r)\over R}\,.
\end{equation}
These models are fully characterized by the initial data,
specified on an initial surface $t=t_{\rm i}$ from which the
collapse develops, which consist of two free functions: the
initial density $\rho_{\rm i}(r)= \rho(t_{\rm i}, r)$ (or
equivalently, the mass function $F(r)$), and $f(r)$, which
describes the initial velocities of collapsing matter shells. At
the onset of collapse the spacetime is singularity-free, so that
by Eq.~(\ref{5}),
\begin{equation}\label{12'}
F(r)=r^3\bar{F}(r)\,,~~~ 0<\bar{F}(0)<\infty \,.
\end{equation}
The initial density $\rho_{\rm i}(r)$ is
\begin{equation}\label{14}
\rho_{\rm i}(r)=r^{-2}F'(r)\,.
\end{equation}

The shell-focusing singularity appears along the curve $t=t_{\rm
s}(r)$ defined by
\begin{equation}\label{s}
R(t_{\rm s}(r),r)=0\,.
\end{equation}
As the density grows without bound, trapped surfaces develop
within the collapsing cloud. These can be traced explicitly via
the outgoing null geodesics, and the equation of the apparent
horizon, $t=t_{\rm ah}(r)$, which marks the boundary of the
trapped region, is given by
\begin{equation}\label{a}
R(t_{\rm ah}(r),r)=F(r)\,.
\end{equation}
If the apparent horizon starts developing earlier than the epoch
of singularity formation, then the event horizon can fully cover
the strong gravity regions including the final singularity, which
will thus be hidden within a black hole. On the other hand, if
trapped surfaces form sufficiently later during the evolution of
collapse, then it is possible for the singularity to communicate
with outside observers.

For the sake of clarity, we consider marginally bound collapse,
$f=0$, although the conclusions can be generalized to hold for the
general case. Then Eq.~(\ref{12}) can be integrated to give
\begin{equation}\label{r}
R^{3/2}(t,r) = r^{3/2} - {\textstyle{3\over2}}(t-t_{\rm i})
F^{1/2}(r)\,,
\end{equation}
and Eqs.~(\ref{s}) and (\ref{a}) lead to
\begin{eqnarray}
t_{\rm s}(r)&=&t_{\rm i}+{2\over3}\left[ {r^3\over
F(r)}\right]^{1/2}\,,\label{s'} \\ t_{\rm ah}(r) &=& t_{\rm s}(r)-
{2\over3}F(r)\,.\label{a'}
\end{eqnarray}
The central singularity at $r=0$ appears at the time
\begin{equation}
t_0=t_{\rm s}(0)=t_{\rm i}+{2\over\sqrt{3\rho_{\rm c}}}\,,
\end{equation}
where where $\rho_{\rm c}=\rho_{\rm i}(0)$. Unlike the homogeneous
dust case (Oppenheimer-Snyder), the collapse is not simultaneous
in comoving coordinates, and the singularity is described by a
curve, the first point being $(t=t_0,r=0)$.

For inhomogeneous dust, Eqs.~(\ref{17}) and (\ref{r}) give
\begin{equation}\label{21}
\sigma^2\equiv {1\over2}\sigma_{ab}\sigma^{ab} = {r\over 6R^4
{R'}^2 F} \left(3F - rF'\right)^2\,.
\end{equation}
A generic (inhomogeneous) mass profile has the form
\begin{equation}\label{19}
F(r)= F_0r^3 + F_1 r^4 + F_2 r^5 +\cdots\,,
\end{equation}
near $r=0$, where $F_0=\rho_{\rm c}/3$. Homogeneous dust
(Oppenheimer-Snyder) collapse has $F_n=0$ for $n>0$, and
Eq.~(\ref{21}) implies $\sigma=0$. The converse is also true in
this case: if we impose vanishing shear $\sigma=0$, we get
$F_n=0$. Whenever there is a negative density gradient, e.g., when
there is higher density at the center, then $F_n\neq0$ for some
$n>0$, and it follows from Eq.~(\ref{21}) that the shear is then
necessarily nonzero. Note that if we want the density profile to
be analytic, we can set all odd terms $F_{2n-1}$ to zero; however,
we note that this is not as such required by our own analysis,
which is independent of any assumptions on $F_n$.

The important question is: what is the effect of such a shear on
the evolution and development of the trapped surfaces? In other
words, we want to determine the behavior of the apparent horizon
in the vicinity of the central singularity at $R=0,r=0$. To this
end, let the first non-vanishing derivative of the density at
$r=0$ be the $n$-th one ($n>0$), i.e.,
\begin{equation}
F(r) = F_0 r^3 + F_n r^{n+3}+\cdots\,,~~ F_n<0\,,
\end{equation}
near the center. By Eqs.~(\ref{21}) and (\ref{a'}),
\begin{eqnarray}
\sigma^2(t,r) &=& {n^2 {F_n}^2 \over 6 F_0}\left[1-3F_0^{1/2}
(t-t_{\rm i}) +{9\over4}F_0 (t-t_{\rm
i})^2\right]r^{2n}\nonumber\\ &&~~~{}+ O(r^{2n+1})  \label{d}\,,
\\ t_{\rm ah}(r) &=& t_0 - {2 \over 3}F_0r^3 - {F_n \over
3{F_0}^{3/2}} r^n+O(r^{n+1})\,.\label{a''}
\end{eqnarray}
The time-dependent factor in square brackets on the right of
Eq.~(\ref{d}) decreases monotonically from 1 at $t=t_{\rm i}$ to 0
at $t=t_0$. Thus the qualitative role of the shear in singularity
formation can be seen by looking at the initial shear. The initial
shear $\sigma_{\rm i}=\sigma(t_{\rm i},r)$ on the surface
$t=t_{\rm i}$ grows as $r^{n}$, $n\geq1$, near $r=0$. A
dimensionless and covariant measure of the shear is the relative
shear, $|\sigma/\Theta|$, where
\begin{equation}
\Theta=2{\dot R\over R}+{\dot R'\over R'}\,,
\end{equation}
is the volume expansion. It follows that
\begin{equation}\label{d'}
\left|{\sigma\over\Theta}\right|_{\rm i}={-nF_n\over3\sqrt{6}
F_0}\, r^n\left[1+O(r)\right] \,.
\end{equation}

It is now possible to see how such an initial shear distribution
determines the growth and evolution of the trapped surfaces, as
prescribed by the apparent horizon curve $t_{\rm ah}(r)$, given by
Eq.~(\ref{a''}). If we assume the initial density profile is
smooth at the center, then $\rho_{\rm i}(r)=\rho_{\rm
c}+\rho_2r^2+\cdots$, with $\rho_2\leq0$, which corresponds to
$F(r)=F_0r^3+F_2r^5+\cdots$, with $F_2\leq0$. Now suppose that
$\rho_2$ (and hence $F_2$) is nonzero. Then Eq.~(\ref{a''})
implies that the apparent horizon curve initiates at $r=0$ at the
epoch $t_0$, and increases near $r=0$ with increasing $r$, moving
to the future. Note that as soon as $F_2$ is nonzero, even with
very small magnitude, the behavior of the apparent horizon changes
qualitatively. Rather than going back into the past from the
center, as would happen in the homogeneous case with $F_2=0$, it
is future pointed. This is what leads to a locally naked
singularity. The singularity may be globally naked, i.e. visible
to faraway observers, depending on the nature of the density
function at large $r$.

A naked singularity occurs when a comoving observer (at fixed $r$)
does not encounter any trapped surfaces until the time of
singularity formation, whereas for a black hole, trapped surfaces
form before the singularity. Thus for a black hole, we require
\begin{equation}\label{bh}
t_{\rm ah}(r)\leq t_0~\mbox{for}~r>0\,,~\mbox{near}~r=0\,.
\end{equation}
In the general case (not necessarily smooth initial density), this
condition is violated for $n=1,2$, as follows from
Eq.~(\ref{a''}). The apparent horizon curve initiates at the
singularity $r=0$ at the epoch $t_0$, and increases with
increasing $r$, moving to the future, i.e. $t_{\rm ah}>t_0$ for
$r>0$ near the center. The behavior of the outgoing families of
null geodesics has been analyzed in detail in these cases, and it
is known that the geodesics terminate at the singularity in the
past~\cite{3}, which results in a naked singularity. In such cases
the extreme strong gravity regions can communicate with outside
observers. For the case $n=3$, Eq.~(\ref{bh}) shows that we can
have a black hole if $F_3\geq -2 F_0^{5/2}$, or a naked
singularity, if $F_3 <-2 F_0^{5/2}$. This is illustrated in
Fig.~1. For $n\geq4$, Eq.~(\ref{bh}) is always satisfied, and a
black hole forms.

\begin{figure}
\includegraphics[height=8cm,width=8cm]{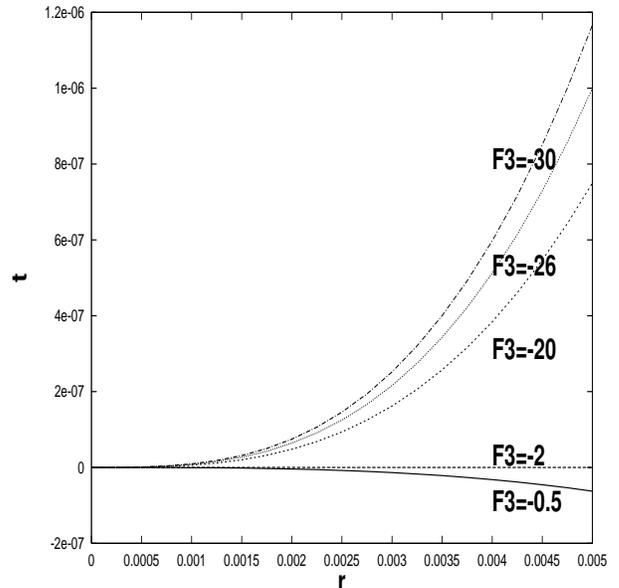}
\caption{ Apparent horizon curves near $r=0$ for the $n=3$ case,
with $F_0=1$. The labels on the curves give the values of $F_3$,
the nonvanishing coefficient quantifying the shear. A black hole
forms if $F_3\geq-2$.}
\end{figure}

When the dust density is homogeneous, the apparent horizon starts
developing earlier than the epoch of singularity formation, which
is then fully hidden within a black hole. There is no density
gradient, and no shear. On the other hand, if a density gradient
is present at the center, then the trapped surface development is
delayed via shear, and, depending on the ``strength" of the
density gradient/shear at the center, this may expose the
singularity. It is the rate of decrease of shear as we approach
the center $r=0$ on the initial surface $t=t_i$, given by
Eq.~(\ref{d'}), that determines the end-state of collapse. {\em
When the shear falls rapidly to zero at the center, the result is
necessarily a black hole; if shear falls more slowly, there is a
naked singularity.} It is thus seen that naked singularities are
caused by the sufficiently strong shearing forces near the
singularity, as generated by the inhomogeneities in density
distribution of the collapsing configuration. When shear decays
rapidly near the singularity, the situation is effectively like
the shear-free (and homogeneous density) case, with a black hole
end-state.

It provides a useful insight to note that when a black hole forms,
the apparent horizon typically springs into being as a
finite-sized surface, at a finite $r$, then moving to the center
$r=0$. This is what happens, for example, in the
Oppenheimer-Snyder black hole formation in homogeneous dust
collapse. In such cases, the event horizon, which does typically
start at a point, could have formed earlier than the apparent
horizon. On the other hand, in the case of a naked singularity, it
follows from Eqs.~(\ref{a'}) and (\ref{a''}), that the apparent
horizon starts at $r=0$, and then is future directed in time, i.e.
$t_{\rm ah}$ grows with increasing coordinate radius $r$ along the
apparent horizon curve $R=F$. These two behaviors of the apparent
horizon curve are very different, and governed by shearing
effects. A comoving observer will {\it not} encounter any trapped
surfaces until the time of singularity formation in the naked
singularity case, whereas in the black hole case, the apparent
horizon typically develops {\it before} the epoch of singularity
formation. This is what we mean by delayed formation of the
apparent horizon, caused by shearing effects.

The relation between density gradients and shear may be understood
via the nonlocal (or free) gravitational field. Density gradients
act as a source for the electric Weyl tensor~\cite{7}
\begin{equation}
{\rm D}^bE_{ab} = {\textstyle{1\over3}}{\rm D}_a\rho\,,
\end{equation}
where ${\rm D}_a$ is the covariant spatial derivative. (The
magnetic Weyl tensor vanishes for spherical symmetry.) In turn,
the gravito-electric field is a source for shear (equivalently,
the shear is a gravito-electric potential~\cite{7}):
\begin{equation}\label{se}
u^c\nabla_c{\sigma}_{ab}+{2\over3} \Theta\sigma_{ab}+
\sigma_{ac}\sigma^c{}_b -{2\over3}\sigma^2 h_{ab}=-E_{ab}\,.
\end{equation}
Thus density gradients may be directly related to shear:
\begin{eqnarray}\label{sd}
{\rm D}_a\rho &=& -4\sigma{\rm D}_a \sigma -2\Theta {\rm
D}^b\sigma_{ab} -3{\rm D}^b\left(u^c\nabla_c
{\sigma}_{ab}\right)\nonumber\\ &&~~{} -3\sigma_a{}^b{\rm D}^c
\sigma_{bc}-3 {\rm D}^b\left(\sigma_{ac}\sigma^c{}_b\right)\,,
\end{eqnarray}
where we have used the shear constraint ${\rm
D}^b\sigma_{ab}={2\over3} {\rm D}_a\Theta$. Equation~(\ref{sd})
makes explicit the link between the behavior of density gradients
and shear near the center, which was discussed above. The free
gravitational field, which mediates this link, can also provide a
covariant characterization of singularity formation. By
Eqs.~(\ref{d'}) and (\ref{se}), the relative gravito-electric
field $E/\Theta^2$ (where $E^2={1\over2} E^{ab}E_{ab}$) near $r=0$
is given at $t=t_{\rm i}$ by
\begin{equation}\label{ee}
\left({E\over\Theta^2}\right)_{\rm i}={-7nF_n\over 18 \sqrt{6}
F_0}\, r^n\left[1+O(r)\right] \,.
\end{equation}
Thus naked singularities in spherical dust collapse are signalled
by a less rapid fall-off of the relative gravito-electric field as
we approach the singularity. Equations~(\ref{d'}) and (\ref{ee})
provide two equivalent ways of expressing the result. This
specifies how much shear is sufficient to create a (locally) naked
singularity.

For the case of dust collapse, the role of shear in deciding the
end-state of collapse is fairly transparent. To understand how
shear affects the formation of the apparent horizon for general
matter fields with pressures included is much more complicated, in
particular since $F=F(t,r)$, whereas $\dot F=0$ for dust. In fact,
even in some general classes of non-dust models (with nonzero
pressure), it is possible to characterize collapse covariantly.
Above we showed that {\em homogeneous density} implies a black
hole end-state. The next logical step would be to consider models
for which the {\em initial} density is homogeneous. For example,
if the mass function is
\begin{equation}\label{mass}
F(t,r)= f(r)-R^3(t,r)\,,~f(r)=2r^3\,,
\end{equation}
then Eq.~(\ref{5}) shows that $\rho_{\rm i}$ and $(p_{\rm r})_{\rm
i}$ are constants. The density and pressure may however develop
inhomogeneities as the collapse proceeds, depending on the choice
of the remaining functions, including in particular the initial
velocities of the collapsing shells, and the collapse may then end
up in either a black hole or a naked singularity, depending on
that (for a discussion on this for the case of dust collapse, we
refer to~\cite{4}). In fact, we can show that {\em zero shear
implies a black hole} for these models. By Eqs.~(\ref{17}),
(\ref{5}) and (\ref{mass}), the shear-free condition leads to
$R'/R=1/r$, and Eq.~(\ref{5}) then shows that $\rho= \rho(t)$,
i.e. the density evolution is necessarily homogeneous. As shown
above, the collapse thus necessarily ends in a black hole. For the
class of models given by Eq.~(\ref{mass}), whenever the collapse
ends in a naked singularity, the shear must necessarily be
nonvanishing. Although this class of models is somewhat special,
the result indicates that the behavior of the shear remains a
crucial factor even when pressures are nonvanishing.

\section{Conclusions}

Since black holes and naked singularities are of great interest in
gravitation theory and astrophysics, it is important to understand
{\it why} these objects develop. The physics of this needs to be
probed carefully in order to make further progress towards cosmic
censorship, or to understand the physical implications of naked
singularities.

It would appear that the only way a singularity can be laid bare
is by distorting the apparent horizon surface and so delay trapped
surface formation suitably. As we have shown here, the shear
provides a rather natural explanation for the occurrence of
(locally) naked singularities. Our main result is that
sufficiently strong shearing effects in spherical collapsing dust
delay the formation of the apparent horizon, thereby exposing the
strong gravity regions to the outside world and leading to a
(locally) naked singularity. When shear decays rapidly near the
singularity, the situation is effectively like the shear-free
case, with a black hole end-state. An important point is that
naked singularities can develop in quite a natural manner, very
much within the standard framework of general relativity, governed
by shearing effects.

In the case of spherical dust collapse, shear and density
inhomogeneity are equivalent, i.e., the one implies the other.
Although shear contributes positively to the focusing effect via
the Raychaudhuri equation,
\begin{equation}
\dot\Theta+{1\over3}\Theta^2=-{1\over 2}\rho-2\sigma^2\,,
\end{equation}
its dynamical action can make the collapse incoherent and
dispersive. (It is this feature which also plays the crucial role
in avoidance of the big-bang singularity in singularity-free
cosmological models~\cite{d}.) {\em Depending on the rate of
fall-off of shear near the singularity, its dispersive effect can
play the critical role of delaying formation of the apparent
horizon, without directly hampering the process of collapse.} The
dispersive effect of shear always tends to delay formation of the
apparent horizon, but is only able to expose the singularity when
the shear is strong enough near the singularity.

We have considered here spherical collapse. Very little is known
about nonspherical collapse, either analytically or numerically,
towards determining the outcome in terms of black holes and naked
singularities. However, phenomena such as trapped surface
formation and apparent horizon are independent of any spacetime
symmetries, and it is also clear that a naked singularity will not
develop in general unless there is a suitable delay of the
apparent horizon. This suggests that the shear will continue to be
pivotal in determining the final fate of gravitational collapse,
independently of any spacetime symmetries. In any case, our main
purpose here has been to try to understand and find the physical
mechanism which leads the collapse to the development of a naked
singularity rather than a black hole in some of the well-known
classes exhibiting such behavior. What we find is that the shear
provides a covariant dynamical explanation of the phenomenon of
naked singularity formation in spherical gravitational collapse.

\newpage
{\bf Acknowledgments:}\\

We thank Shrirang Deshingkar for help with the figure. This work
arose out of discussions when PSJ and ND were visiting the
University of Natal, and they thank Sunil Maharaj for discussions
and warm hospitality. Part of the work was done during visits by
RM and ND to the Tata Institute, Mumbai, by PSJ and RM to IUCAA,
Pune, and by ND to the University of Portsmouth (supported by
PPARC).


\begin{references}


\bibitem{1}
For recent reviews, see, e.g., P. S. Joshi, Pramana {\bf 55}, 529
(2000); C. Gundlach, Living Rev. Rel. {\bf 2}, 4 (1999); A.
Krolak, Prog. Theor. Phys. Suppl. {\bf 136}, 45 (1999); R.
Penrose, in {\it Black holes and relativistic stars}, ed. R. M.
Wald (University of Chicago Press, 1998); T. P. Singh,
gr-qc/9805066.

\bibitem{jdm}
T. Harada, H. Iguchi, and K.I. Nakao, Phys. Rev. D {\bf 61},
101502 (2000);  P. S. Joshi, N. Dadhich, and R. Maartens, Mod.
Phys. Lett. {\bf A15}, 991 (2000); T. Harada, H. Iguchi, K.I.
Nakao, T. P. Singh, T. Tanaka, and C. Vaz, Phys. Rev. D {\bf 64},
041501 (2001).

\bibitem{hm}
For recent examples, see, e.g., H. Iguchi, K.I. Nakao, and T.
Harada, Phys. Rev. D {\bf 57}, 7262 (1998); S. S. Deshingkar, P.
S. Joshi, and I. H. Dwivedi, Phys. Rev. D {\bf 59}, 044018 (1999);
R. V. Saraykar and S. H. Ghate, Class. Quantum Grav. {\bf 16}, 281
(1999); B. J. Carr and A. Coley, Class. Quantum Grav. {\bf 16},
R31 (1999); S. M. Wagh and S. D. Maharaj, Gen. Rel. Grav. {\bf
31}, 975 (1999); B.C. Nolan, Phys. Rev. D {\bf 60}, 024014 (1999);
F. C. Mena, R. Tavakol, and P. S. Joshi, Phys. Rev. D {\bf 62},
044001 (2000); S. Jhingan and G. Magli, Phys. Rev. D {\bf 61},
124006 (2000); S. M. C. V. Goncalves, Phys. Rev. D {\bf 63},
124017 (2001); S. Jhingan, N. Dadhich, and P. S. Joshi, Phys. Rev.
D {\bf 63}, 044010 (2001); T. Harada and H. Maeda, Phys. Rev. D
{\bf 63}, 084022 (2001); S. M. C. V. Goncalves and S. Jhingan,
gr-qc/0107054; F. Mena and B.C. Nolan, Class. Quantum Grav. {\bf
18}, 4531 (2001) ; S. G. Ghosh and A. Beesham, Phys. Rev. D {\bf
64}, 124005 (2001).

\bibitem{3}
P. S. Joshi and I. H. Dwivedi, Class. Quantum Grav. {\bf 16}, 41
(1999).

\bibitem{4}
P. S. Joshi and I. H. Dwivedi, Phys. Rev. D {\bf 47}, 5357 (1993).

\bibitem{7}
R. Maartens and B. A. Bassett, Class. Quantum Grav. {\bf 15}, 705
(1998).

\bibitem{d}
N. Dadhich, J. Astrophys. Astr. {\bf 18}, 343 (1997); N. Dadhich
and A. K. Raychaudhuri, Mod. Phys. Lett. {\bf A14}, 2135 (1999).


\end{references}
\end{document}